\documentstyle[12pt]{article}
\input{psfig}
\begin{document}

\author{Yu.M. Sinyukov, S.V. Akkelin \\
Bogolyubov Institute for Theoretical Physics,\\
Kiev 252143 , Ukraine \and Nu Xu \\
Lawrence Berkeley National Laboratory,\\
Berkeley, CA 94720, USA}
\title{On final conditions in high energy heavy ion collisions }
\date{\today}
\maketitle

\begin{abstract}
Motivated by the recent experimental observations, we discuss the
freeze-out properties of the fireball created in central heavy ion
collisions.  We find that the freeze-out conditions, like temperature,
velocity gradient near center of the fireball, are similar for
different colliding systems and beam energies. This means that the
transverse flow is stronger in the collisions of heavy nuclei than
that of the light ones. 
\end{abstract}

The system that created in relativistic heavy-ion collision can have
both longitudinal and transverse expansion (see, e.g.,
\cite{heinz}). In order to study the hadronic experimental data
\cite{pratt98,nix98}, one needs a mathematical description of a final
stage of collisions. At the region near mid-rapidity the
boundary effect, due to the finite length of the hydrodynamic tube in
longitudinal direction, can be neglected \cite{yuri1}. It is possible
to use Bjorken's model for longitudinal expansion: $v_l=x_l/t$. The
same quasi-inertial flow is inherent to Landau model at freeze-out
stage \cite {landau,bjorken}. In this approach, the parameter $\tau$
was introduced to describe proper time of the expanding system. For
the transverse expansion, we will use a rather general picture
proposed in Ref.\cite{yuri2} where, due to the cylindrical symmetry,
one has $v_T(0)=0$ with the derivative of the transverse velocity near
the center of the fireball $v_T^{\prime }(0)\neq 0$. The transverse
velocity increases monotonously as a function of radius $r$.

 The single particle spectra in a pure hydrodynamic picture without
resonance decays are expressed by the integral of the Wigner function
over the freeze-out surface:

\begin{equation}
p^0\frac{d^3N}{d^3p}=\int d\sigma _\mu p^\mu f(x,p)  \label{spectra-def}
\end{equation}

 The Wigner function $f(x,p)=f_{th}(x,p)\cdot \rho (x)$, where
$f_{th}(x,p)$ is the local thermal distribution function with a
temperature parameter $T=1/\beta $, chemical potential $\mu $ and
4-velocity field $u(x)$:

\begin{equation}
f_{th}(x,p)=\frac{(2j+1)}{(2\pi )^3}\frac 1{\exp \left( \beta p^\mu u_\mu
(x)-\beta \mu \right) \mp 1}  \label{Wigner-therm}
\end{equation}

 The function 

\begin{equation}
\rho (x)\propto \exp \left[ \frac \alpha 2\left( u(r,y_L)-u(0,y_L)\right)
^2\right] =\exp \left[ -\alpha \left( \cosh y_T(r)-1\right) \right]
\label{ro}
\end{equation}

{\noindent is introduced to describe the finiteness of the
relativistically expanding system in transverse direction
\cite{yuri2}. Note that the form of the transverse rapidity $y_T$
dependent density is similar to one used in Ref. \cite{satz}. Here
$y_L$ and $y_T$ are the longitudinal and transverse rapidities
respectively and $\alpha $ is the intensity of transverse flows }

\begin{equation}
\alpha =(v_T^{\prime }(0)\cdot \overline{R}_T)^{-2}.  \label{alph-def}
\end{equation}
For very intensive relativistic transverse flows, $\alpha \rightarrow 0$. On
the other hand for $\alpha \gg 1$ we have non-relativistic transverse flows
and
\begin{equation}
\rho (x)\propto \exp (-r^2/2\overline{R}_T^2)  \label{ro1}
\end{equation}
for $y_T(r)\ll 1$.

Using the saddle point approximation , we have obtained from Eq. (\ref
{spectra-def}), for $m_T\beta \gg 1$,  $m_T=\sqrt{m^2+p_T^2}$ \cite{yuri2}:

\begin{equation}
\frac{d^2N}{m_Tdm_Tdy}\propto e^{-(\beta m_T+\alpha )(1-\overline{v}%
_T^2)^{1/2}}  \label{spectra-approx}
\end{equation}
\noindent where transverse velocity at the saddle-point $\overline{r}(p)$ is

\begin{equation}
\overline{v}_T\equiv \tanh y_T(\overline{r}(p))=\frac{\beta p_T}{\beta
m_T+\alpha }  \label{velocity}
\end{equation}

 Within the model described in Eqs.(1)-(3), the transverse momentum
distribution (6) is not very sensitive to the details of the velocity
profile, see Ref. [5]. In the region where

\begin{equation}
m_T-m\ll \frac{(\beta m+\alpha )^2}{2\beta \alpha }  \label{Teff-condition}
\end{equation}
one has the relationship:

\begin{equation}  \label{relationship}
(\beta m_T+\alpha )(1-\overline{v}_T^2)^{1/2}\cong (\beta m+\alpha )+\frac{%
\alpha \beta (m_T-m)}{(\beta m+\alpha )}
\end{equation}
Therefore we have the simplified Eq. (\ref{spectra-approx}) as:

\begin{equation}
\frac{d^2N}{m_Tdm_Tdy}\sim e^{-\beta _{eff}(m_T-m)}  \label{spectra-eff}
\end{equation}
with\footnote{ The linear dependence of the slope parameter of the
transverse momentum distributions on particle mass was obtained early
for spherically symmetric non-relativistic expanding system in
Ref. \cite {Csorgo}.}

\begin{equation}
1/\beta _{eff}=T_{eff}=T+\frac m\alpha.  \label{Teff-def}
\end{equation}
The latter equation connects the slope of the transverse mass spectra with
the intensity of transverse flows
\begin{equation}
\alpha ^{1/2}=\frac{\mbox{hydrodynamical length}}{\mbox{transverse radius}}.
\label{alpha}
\end{equation}
The intensity of transverse flows $\alpha $ is defined by the
Eq. (\ref {alph-def}) and does not depend on $m$, it depends only on
$A$, where $A$ is the nucleus atomic number. 

The system finally decays when the rarefaction wave reaches the center
part of the fireball. The analysis of the experimentally measured
slope parameters ($T_{eff}$) as a function of particle mass
\cite{na44slope} with two parameters, $T$ and $\alpha ,$ shows that
the freeze-out temperature $T$ is approximately constant for different
colliding systems at the beam energy $\geq$ 10A$\cdot$GeV. Hence it is
natural to conclude that the physical condition near the center is the
same for different colliding systems, i.e. $v_T^{\prime }(0)$ is
constant. From Eq.(\ref{alph-def}) we obtain

\begin{equation}
v_T^{\prime }(0)=(\alpha \overline{R}_T^2)^{-1/2}
\label{velocity-derivative}
\end{equation}

 In our approach $\overline{R}_T$ is the Gaussian-like radius of a
 decaying system. We suppose that $\overline{R}_T$ depends on $A$
 only. It means that in the same collisions a freeze-out ``size'' is
 unique for different particle species. The value $\overline{R}_T$ is
 connected with the so called sideward radius $R_s$ \cite{pratt86},
 obtained from the fit of the experimentally measured two-particle
 correlation functions, in the following way.  When
 $\overline{v}_T=\frac{\beta p_T}{\beta m_T+\alpha }\ll$ 1
 (non-relativistic approximation) it can be shown that only
 $v_T^{\prime }(0)$ has an influence on the interferometry radii and
 any details of the transverse velocity profile $v_T(r)=\tanh y_T(r)$
 are not important \cite{yuri2,yuri3}. \footnote{Assuming the validity
 of this approach, we found that, even for the heaviest colliding
 system Pb+Pb [10], $\alpha = 6.57$ and $\overline{v}_T$ = 0.32. These
 values were obtained at $m_T \approx 0.5$ GeV where is the measured
 highest pair momentum region. For smaller $p_T$ and lighter colliding
 system, this assumption works better.}  Within this approximation,
 one has the following expression for the Gaussian transversal
 sideward radius \cite{yuri2,yuri3}:

\begin{equation}
\overline{R}_T\sqrt{\frac \alpha {\beta m_T+\alpha }}=R_S(p_T).
\label{radius-trans}
\end{equation}

 From this equation we evaluate $\overline{R}_T$, using experimental
 data for $R_S(p_T)$, and $\beta $ and $\alpha $ from the fitting
 procedure \cite {na44slope} based on Eq. (\ref{Teff-def}). To
 minimize the influence of the resonance decays, the interferometry
 radius for every analyzed particle species has to be measured at the
 point where $p_T$ is large enough. On the other hand, in order to
 provide a validity of the non-relativistic approximation and the
 correctness of the condition Eq. (\ref{Teff-condition}), the value of
 $p_T$ should be limited with $\overline{v}_T \ll 1$.

 Using the experimental data, listed in Table I, we evaluate the
 freeze-out temperature and velocity gradient at the center of the
 fireball. The results of the fit are shown in Fig. 1. Note that there
 are two important features in the plot. First, the values of the
 intrinsic temperature $T$ seems to be a constant (top plot of
 Fig. 1). It does not dependent on the size of colliding system.
 Furthermore, for collisions at the AGS energies (11 - 15 A$\cdot
 $GeV/c) and SPS energies (158 - 200 A$\cdot $GeV/c), this parameter
 is approximately the same. Secondly, the value $v_T^{^{\prime }}(0)$
 is also closed to a constant (see bottom plot of Fig. 1).

 It has been noted that the temperature is limited for collisions at
 beam energies larger than 11 A$\cdot $GeV (Fig. 2 of Ref. \cite
 {na44slope}). One should note that, although we do not observe the
 difference in $v_T^{^{\prime }}(0)$ between the collisions at AGS and
 SPS energies, it is quite possible that at lower beam energies one
 gets somewhat different values. The saturation of the freeze-out
 temperature might reflect the limiting temperature hypothesis pointed
 by Hagedorn some years ago \cite {heg,stocker}. On the other hand,
 the constant behavior of the velocity gradient near the center of the
 fireball is a new result yield by the present study. It clearly
 shows that at the final stage, the freeze-out conditions are
 approximately the same for different size and different beam
 energies of the collisions.  Constant transverse velocity gradient
 actually means that both the averaged transverse velocity and flow
 intensity grow with the size of the colliding nuclei. Indeed, the
 average transverse hydrodynamic velocity that can be calculated using
 the final distribution function (\ref{Wigner-therm}) is

\begin{equation}
\left\langle v_T\right\rangle =\frac{\int \frac{d^3p}{p^0}%
\int d\sigma _\mu p^\mu v_T(x)f(x,p)}{\int \frac{d^3p}{p^0}\int d\sigma _\mu
p^\mu f(x,p)}=\frac{\int d\sigma _\mu u^\mu (x)\rho (x)v_T(x)}{\int d\sigma
_\mu u^\mu (x)\rho (x)}  \label{vt}
\end{equation}
For $\tau (r)\cong const$ at the freeze-out hypersurface we can get from Eq.
(\ref{vt}): 
\begin{equation}
\left\langle v_T\right\rangle \cong \frac{\int \tanh y_T(r)\cosh
y_T(r)\exp \left[ -\alpha \cosh y_T(r)\right] rdr}{\int \cosh y_T(r)\exp
\left[ -\alpha \cosh y_T(r)\right] rdr}.  \label{vt1}
\end{equation}
Using saddle point approximation, in the limit of $\alpha \gg 1$, one
can show that the average transverse hydrodynamic velocity is
\begin{eqnarray}
\left\langle v_T\right\rangle &\cong &v_T^{\prime }(0)\frac{\int
\exp (-r^2/2\overline{R}_T^2)r^2dr}{\int \exp (-r^2/2\overline{R}_T^2)rdr}%
\cong   \nonumber  \\
&&  \label{vt2} \\
&\cong &\sqrt{\frac \pi 2}v_T^{\prime }(0)\overline{R}_T=\sqrt{\frac \pi
{2\alpha }}  \nonumber
\end{eqnarray}

 Let us note that $\overline{v}_T\neq \left\langle v_T\right\rangle $.
 The physical meaning is that, in the saddle point approximation,
 $\overline{v}_T$ is the transverse velocity of the fluid element
 which gives the main contribution to transverse momentum spectrum at
 given $p_T$.  On the other hand, $\left\langle v_T\right\rangle $ is
 the averaged transverse hydrodynamic velocity that characterizes whole
 system. From Eqs. (16) and (17), one can see that $\alpha $ is directly
 connected with $ \left\langle v_T\right\rangle $.  So we find that,
 in approximation of Eq. (17), the averaged values of the transverse
 velocity $\langle v_T \rangle$ are 0.49$c$ and 0.33$c$ for Pb+Pb and
 S+S central collisions [10], respectively. Formally the averaged
 velocity $\langle v_T \rangle$ is consistent with zero for p+p
 collisions although the application of the thermal model for such
 collisions remains an open question. At the freeze-out surface, the
 effective slopes of spectra for different colliding systems and
 different mass particle species can be presented by the
 non-relativistic average transverse hydrodynamic velocity:

\begin{equation}
T_{eff} \approx T+\frac 2\pi m\left\langle v_T\right\rangle ^2  \label{Teff}.
\end{equation}

 The merit of this equation is that by fitting the measured slope
 parameter as a function of particle mass, one can separate the
 collective motion from the thermal motion. Therefore, the intrinsic
 freeze-out temperature $T$ and averaged collective velocity $\langle
 v_T \rangle$ can be readily extracted from the data. The measured
 slope parameters of pions, kaons, and protons [10] seem to obey the
 linear mass dependence as given by Eq. (18). One new finding from
 this study is the linear tie of the averaged transverse velocity
 $\langle v_T \rangle$ with the radius $\overline{R}_T$, see
 Eq. (17).  It is worth mentioning that the hydrodynamic arguments for
 the behavior of the slope parameter as a function of a particle mass
 is valid as long as the considered particle species remain a part of
 the fireball and hence are participating in the frequent
 rescatterings. For those particles with high probability of
 destruction, such description is not valid anymore. It is possible
 that those particles freeze-out earlier than the other hadrons which
 participate in the evolution of the system longer. They do not have
 enough time to acquire a common collective velocity and hence the
 corresponding slope parameters could be smaller than the predicted by
 Eq. (18). Indeed, the recent preliminary results reported by WA97
 \cite{wa97} and NA49 \cite{na49} show such deviations for hyperons,
 particularly for $\Omega$ particle \cite{hsx}. In order to understand
 the dynamics involved for those particles, one has to take the
 collision rate for individual considered particle into account.

 In summary, using a thermal model we analyzed the recent experimental
data from heavy ion collisions. We found that the freeze-out
temperature is a constant for all colliding systems at $E_{beam}
>$10 A$\cdot$GeV and a constant velocity gradient at the center of the
fireball ($r \rightarrow 0$).

\vspace{0.2in}

{\bf Acknowledgments}: One of us (Nu Xu) wishes to thank Professor H.
St\"{o}cker for many helpful discussions on the physics of
thermalization and hydrodynamic flow in heavy-ion collisions. We thank
Dr. S. Panitkin for careful reading of the manuscript. We gratefully
acknowledges support by the Ukrainian State Fund of the Fundamental
Research under Contract No 2.5.1/057 and by the U.S. Department of
Energy under Contract No. DE-AC03-76SF00098.


\begin{table}
\begin{tabular}{||c||c|c||} \hline\hline
 System       & 1/$\sqrt{\alpha}$  & $R_S$         \\ \hline 
 p + p [10]   & 0.01$\pm$0.11      & 0.8$\pm$0.2   \\ \hline
 S + S [10]   & 0.26$\pm$0.10      & 2.9$\pm$0.2   \\ \hline
Pb + Pb [10]  & 0.39$\pm$0.12      & 5.1$\pm$0.4   \\ \hline
Si + Al [13]  & 0.29$\pm$0.12      & 2.56$\pm$0.17 \\ \hline
Si + Au [13]  & 0.30$\pm$0.12      & 2.4$\pm$1.1   \\ \hline
Au + Au [14]  & 0.42$\pm$0.11      & 3.57$\pm$0.52 \\ \hline\hline
\end{tabular}
\caption{Experimental transverse velocity intensity and sideward
radius parameters for different colliding systems. Error bars are
statistical only. Note that for the AGS energies, the size parameters
$R_S$ is taken from 2-dimensional fits assuming that $R_T = R_S$. All
size parameters are obtained with a cut $p_T >$ 0.2 GeV/c.}
\end{table}

\begin{figure}
\psfig{file=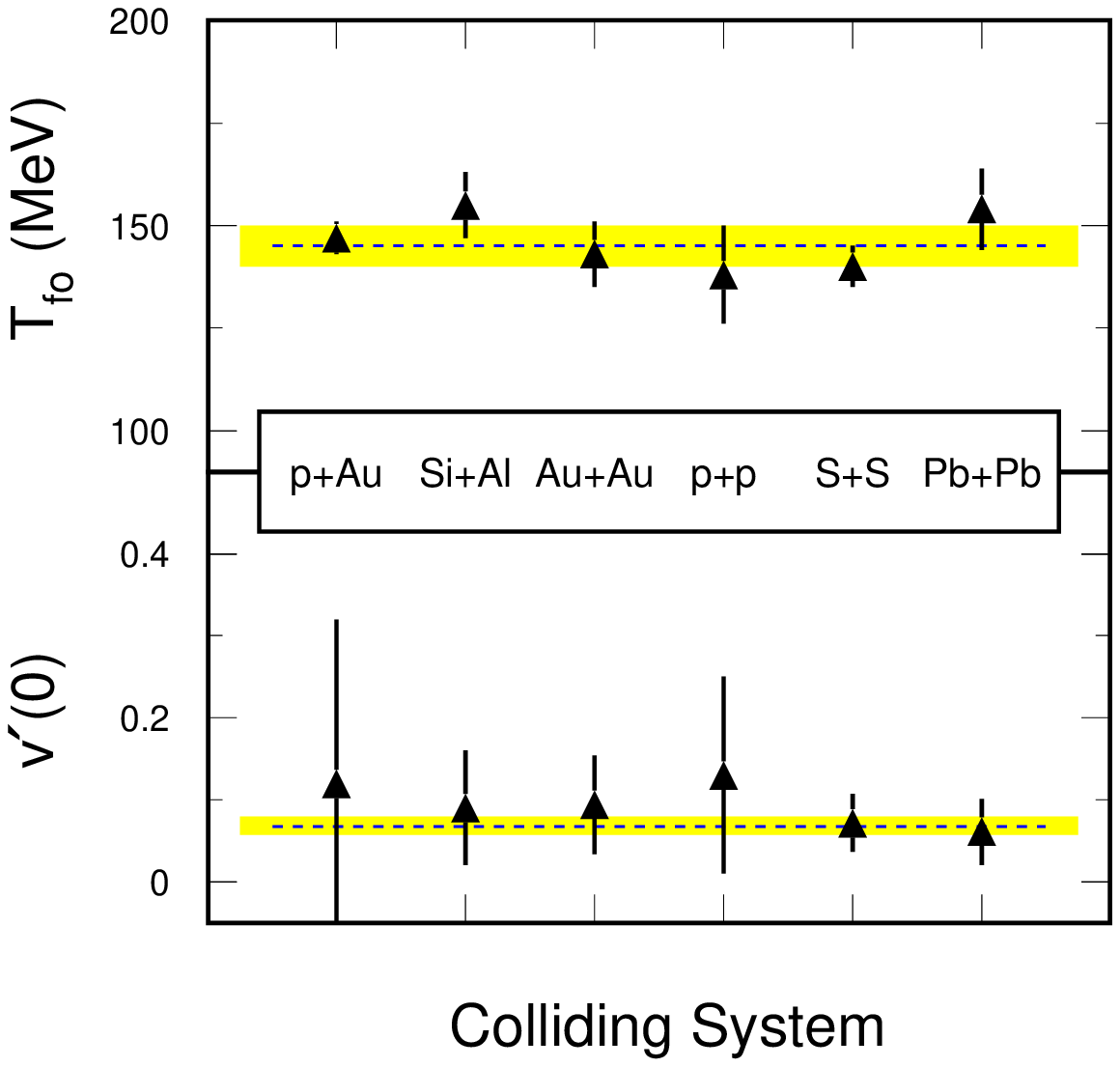,width=14.0cm}
\caption{ The freeze-out temperature $T$ (top) and flow velocity
gradient (bottom) calculated at $r \approx$ 0. The set of data for this
calculation is the taken from the references listed in Table I.}
\end{figure}


\begin{thebibliography}{10}
\bibitem{heinz}  E. Schnedermann, J. Sollfrank, and U. Heinz, Phys. Rev.
C48, 2462 (1993).

\bibitem{pratt98} S. Pratt and J. Murray, Phys. Rev. {\bf C57},
1907(1998).

\bibitem{nix98} J.R. Nix {\it et al.}, in {\it Advances in Nuclear
Dynamics 4}, Proc. 14th Winter Workshop on Nuclear Dynamics, Snowbird,
Utah, 1998 (Plenum Press, New York, 1998).

\bibitem{yuri1}  V.A. Averchenkov, A.N. Makhlin, Yu.M. Sinyukov,
Sov.J.Nucl.Phys. {\bf 46}, 905 (1987).

\bibitem{landau}  L.D. Landau, Izv. Akad. Nauk SSSR Ser. Fiz. 17, 51
(1953).

\bibitem{bjorken}  J.D. Bjorken, Phys. Rev. {\bf D27}, 140 (1983).

\bibitem{yuri2} S.V. Akkelin and Yu.M. Sinyukov, Phys. Lett. {\bf
B356}, 525 (1995); S.V. Akkelin and Yu.M. Sinyukov, Z. Phys. {\bf
C72}, 501 (1996).

\bibitem{satz}  A. Leonidov, M. Nardi and H. Satz, Nucl. Phys. A610, 124c
(1996).

\bibitem{Csorgo}  T. Csorgo, B. Lorstad, J. Zimanyi, Phys. Lett. B338,
134 (1994).

\bibitem{na44slope}  I.G. Bearden {\it et al}., (The NA44 Collaboration),
Phys. Rev. Lett. 78, 2080 (1996).

\bibitem{pratt86} S. Pratt, Phys. Rev. {\bf D33}, 1314(1986).

\bibitem{yuri3}  Yu.M. Sinyukov, S.V. Akkelin, A.Yu. Tolstykh, Nucl.
Phys. A610, 278c (1996).

\bibitem{cianciolo95} V. Cianciolo, Nucl. Phys. {\bf A590}, 459c(1995).

\bibitem{baker96} M.D. Baker, Nucl. Phys. {\bf A610}, 213c(1996).

\bibitem{heg}  R. Hagedorn, Suppl. Nuovo Cim. {\bf III.2}, 147 (1965).

\bibitem{stocker}  H. St\"{o}cker, A.A. Ogloblin, and W. Greiner, Z.
Phys. {\bf A303}, 259 (1981).

\bibitem{wa97} I. Kralik {\it et al}., (The WA97 Collaboration), Quark
Matter '97, Tsukuba, Japan, Dec. 1 - 5, 1997.

\bibitem{na49} G. Roland {\it et al}., (The NA49 Collaboration), Quark
Matter '97, Tsukuba, Japan, Dec. 1 - 5, 1997.

\bibitem{hsx} H. van Hecke, H. Sorge, N. Xu, submitted to
Phys. Rev. Lett., April 1998.

\end{thebibliography}
\end{document}